\documentstyle[e-e-ijmpa,twoside]{article}
   \input epsf

\renewcommand{\thefootnote}{\fnsymbol{footnote}}

\def\beq{\begin{equation}}
\def\eeq#1{\label{#1}\end{equation}}
\def\eeqn{\end{equation}}
\def\beqa{\begin{eqnarray}}
\def\eeqa#1{\label{#1}\end{eqnarray}}
\def\eeqan{\end{eqnarray}}
\def\CR{\nonumber \\ }
\def\leqn#1{(\ref{#1})}

\def\Journal#1#2#3#4{{#1} {\bf #2}, #3 (#4)}

\def\NPB{{\em Nucl. Phys.} B}

\def\PRD{{\em Phys. Rev.} D}

\def\PRT{{\em Phys. Repts.}}
\def\IJMPA{{\em Int. J. Mod. Phys.} A}

\let\bar=\overbar
\def\Dslash{\not{\hbox{\kern-4pt $D$}}}
\def\dslash{\not{\hbox{\kern-2pt $\del$}}}
\def\half{\frac{1}{2}}

\def\del{\partial}

\def\ee{e^+e^-}
\def\emem{e^-e^-}
\def\se{\widetilde{e}}

\def\mz{m_Z}

\def\msb{{\bar{\ssstyle M \kern -1pt S}}}

\def\ELER{e^-_Le^+_R}
\def\EREL{e^-_Re^+_L}
\def\ELEL{e^-_Le^+_L}
\def\ERER{e^-_Re^+_R}
\def\EELL{e^-_Le^-_L}
\def\EELR{e^-_Le^-_R}
\def\EERR{e^-_Re^-_R}
\def\SE#1#2{\se^-_{#1} \se^+_{#2}}
\def\SEE#1#2{\se^-_{#1} \se^-_{#2}}
\def\ch#1{\widetilde\chi^+_{#1}}

\def\ne#1{\widetilde\chi^0_{#1}}

\newcommand\pubnumber{\large SLAC-PUB-7759}
\newcommand\pubdate{\large March, 1998}

\def\Title#1{\begin{center} {\Large #1 } \end{center}}
\def\Author#1{\begin{center}{ \sc #1} \end{center}}
\def\Address#1{\begin{center}{ \it #1} \end{center}}

\def\doeack{\footnote{Work supported by the Department of Energy,
                     contract DE--AC03--76SF00515.}}
\def\SLAC{Stanford Linear Accelerator Center\\
    Stanford University, Stanford, California 94309 USA}

\newenvironment{Abstract}{\begin{quotation} \begin{center}
                       ABSTRACT
     \end{center}\bigskip  }{\end{quotation}}
\begin{document}
\begin{flushright}\begin{tabular}{l} \pubnumber\\
         \pubdate  \end{tabular}\end{flushright}
\vfill
\Title{Systematics of Slepton Production in $\ee$ and $\emem$ Collisions}
\vfill
\Author{Michael E. Peskin\doeack}
\Address{\SLAC}

\vfill
\begin{Abstract}
I present the basic formulae for slepton production 
in $\ee$ and $\emem$ collisions in an especially simple form, using a 
helicity basis.  This parametrization introduces the useful 
{\em neutralino functions} to connect the neutralino eigenstates 
to observable cross sections.
\end{Abstract}
\vfill

\vfill

\begin{center}
  to appear in the proceedings of the\\  Second International Workshop on
     Electron-Electron Interactions at TeV Energies\\
     Santa Cruz, California, 22-24 September 1997
\end{center}
\vfill
\normalsize\textlineskip

\newpage

\hbox to \hsize{\null}
\setcounter{page}{0}
\newpage

\title{SYSTEMATICS OF SLEPTON PRODUCTION IN {\boldmath $\ee$} and
{\boldmath $\emem$} COLLISIONS}

\author{MICHAEL E. PESKIN%
\footnote{Work supported by the Department of Energy, 
Contract  DE--AC03--76SF00515.}
}
\address{Stanford Linear Accelerator Center \\
 Stanford University, Stanford, California 94309 USA}

\maketitle\abstracts{I present the basic formulae for slepton production 
in $\ee$ and $\emem$ collisions in an especially simple form, using a 
helicity basis.  This parametrization introduces the useful 
{\em neutralino functions} to connect the neutralino eigenstates 
to observable cross sections.}

\setcounter{footnote}{0}
\renewcommand{\thefootnote}{\alph{footnote}}

\vspace*{1pt}\textlineskip	

\section{Introduction}

Today, it seems that supersymmetry is the extension of the Standard Model
most likely to be observed at high energies.  Thus, when we consider any
future accelerator, it is important to pay attention to its capabilities 
for studies of supersymmetry.  By this, I mean not only the first
discovery of 
supersymmetric particles but also the systematic measurement of the 
mass spectrum of supersymmetric particles.  If supersymmetry is indeed
a property of Nature, the supersymmetry spectrum can be a window into 
physics at very small distances, perhaps even into the most fundamental 
interactions.\cite{MEPMo,Mura}
 To look through this window and see everything that it presents
to us, we will need a variety of effective tools.

 The $\emem$ collider can access supersymmetric
particles through the reactions
\beq 
         \emem \to \se^- \se^- \ ,
\eeq{basic}
with the partners of the left- and right-handed electron and positron 
 in the final 
state.  As long as the selectrons are light enough to be pair-produced, 
the cross sections for these processes are substantial, of the order
of a unit of $R$.  These processes lead to characteristic final states
consisting of $\emem$ plus missing neutral particles. 

 Cuypers, van 
Oldenbourgh, and R\"uckl\cite{CVR,CMor,CEE} have studied the observability
of this signature in some detail.  They have found that it is rather 
easy to detect even in the region of parameters $\mu \ll m_2$ which presents
a special problem for supersymmetry searches.  The irreducible Standard Model
background $\emem \to \emem Z^0$ can be eliminated using the final state
kinematics, and the other dominant background,
from $\emem\to e^- \nu W^-$, with the $W^-$
decaying leptonically and one electron lost, can be controlled by kinematic
cuts and the use of initial-state polarization.  Thus, the reaction
 \leqn{basic} could provide an especially clean sample of events for 
precision studies.

The reactions \leqn{basic} procede by exchange of a neutralino.  Thus, they
potentially give information on the spectrum and mixing angles of the 
neutralinos.  The information one obtains is similar to that provided by 
the reactions
 \beq 
         \ee \to \se^+ \se^- \ .
\eeq{basicep}
 Though these reactions can also involve $s$-channel annihilation through
 a photon or a $Z^0$, typically the $t$-channel neutralino exchange diagrams
are dominant.

  To compare $\ee$ and $\emem$ reactions more carefully, it is 
useful to present the formulae for these cross sections as simply as possible.
Of course, these cross sections are well known; they can be found, for
example, in the compendium of Baer, Bartl, Karatas, Majerotto, and 
Tata,\cite{BBKMT} and some are even given in the famous review of 
Haber and Kane.\cite{HK}  The sensitivity of slepton production to the 
neutralino parameters have been studied in  detailed simulations 
by the JLC group.\cite{Tsuk,FNT}  Still, it never hurts to make the 
basic interrelations more transparent.
  That is the goal of this short paper. 

\section{Neutralino Mixing}

To begin, let me review a bit of the formalism of neutralino mixing.
The neutralino mass matrix, written in the basis of superpartners of the 
$U(1)$ and neutral $SU(2)$ gauge bosons and the Higgsinos $(\widetilde b,
\widetilde w^3, i\widetilde h_1^0, i\widetilde h_2^0)$, takes the form
\beq
 \mbox{\bf m} = \pmatrix{m_1 & 0 &  - \mz s_w \cos\beta & \mz s_w\sin\beta \cr
             0 & m_2 &  \mz c_w \cos\beta & - \mz c_w\sin\beta \cr
             - \mz s_w \cos\beta & \mz c_w\cos\beta & 0 & -\mu\cr
             \mz s_w \sin\beta & - \mz c_w\sin\beta &  -\mu & 0 \cr} \ ,
\eeq{nmat}
with $(c_w,s_w)= (\cos\theta_w, \sin\theta_w)$.
This matrix depends on paramters $m_1$, $m_2$, $\mu$, of which $m_2$
can be chosen to be positive. 
In grand unified or gauge-mediated models of supersymmetry-breaking,
$m_1 = 0.5 m_2$, and I will assume this in computing the 
 curves displayed in this paper.
However, it is important to state that this ratio could have a wide
range of values and must ultimately be determined experimentally.

\begin{figure}
\begin{center}
\leavevmode
{\epsfxsize=3.20truein \epsfbox{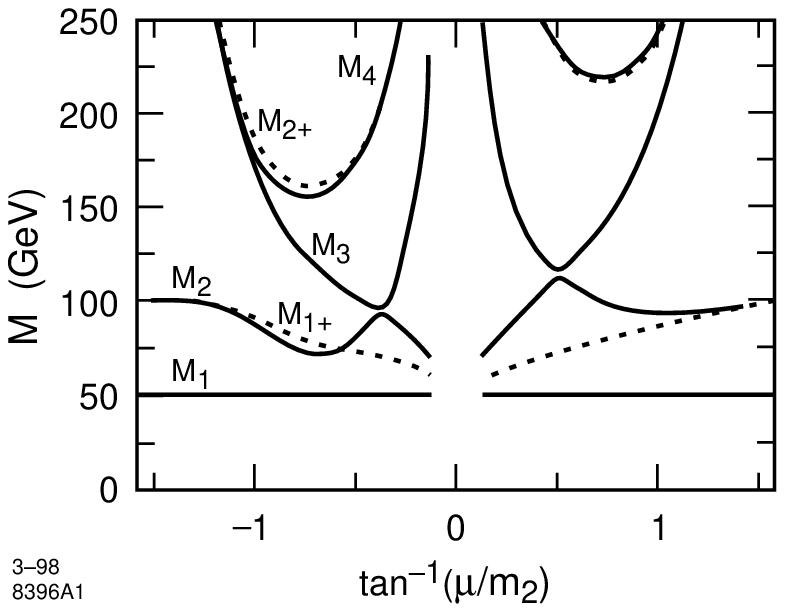}}
\end{center}
 \fcaption{The spectrum of neutralinos displayed as a function of 
       $\tan(\mu/m_2)$, for $\tan \beta=4$,
along a line in parameter space for which
   the mass of the lightest neutralino is 50 GeV.}
\label{Nmass}
\end{figure}

The eigenstates of the matrix \leqn{nmat}
change qualitatively depending on whether
$m_2 > |\mu|$ or $m_2 < |\mu|$.  In the first case, the lightest
neutralino is approximately the $\widetilde{b}$.  We might call this the
`gaugino' region.  In the second case, the lightest neutralino is a 
linear combination of the two Higgsinos.  This is the `Higgsino' region.
The crossover between these two regions is illustrated in Figure \ref{Nmass},
in which I show the dependence of the masses of the four neutralinos
$\ne{i}$ 
and also of the two charginos $\ch{i}$ on the ratio $m^2/\mu$, for
parameters that keep the mass of the lightest neutralino fixed.

Typically, the whole pattern of decays of supersymmetric particles is
affected by the value of $m_2/\mu$, so this parameter must be determined
accurately to carry out any precision studies.  We will see that the
various processes \leqn{basic} and \leqn{basicep} are nicely sensitive
to $m_2/\mu$ and $m_1/m_2$.

The eigenstates of the mass matrix \leqn{nmat} enter the theory of $\ee$
and $\emem$ scattering because the gaugino components of the neutralino
mediate the transitions from electron to selectron states.  Write the 
diagonalization of the matrix \leqn{nmat} as
\beq
   \mbox{\bf m} =  V  D  V^\dagger   \ ,
\eeq{diagV}
where  $D$ is a diagonal matrix whose eigenvalues are $M_i$, $i=1,\ldots,4$,
arranged in ascending order by absolute value.  Then, for example, $V_{1i}$
gives the $\widetilde b$ admixture in the neutralino $i$.  To present
formulae for the $\ee$ and $\emem$ processes of interest, let
\beq
   V_{Li} = {1\over 2c_w} V_{1i} + {1\over 2s_w} V_{2i} \ ,\qquad 
     V_{Ri} =  {1\over c_w} V_{1i} \ .
\eeq{Vdefs}
From these objects, we can construct the  dimensionless
{\em neutralino functions},
\beqa
      N_{ab}(t) &=& \sum_i   \, V_{ai} {M_1^2 \over M_i^2 - t} V_{bi}^* \CR
    M_{ab}(t) &=& \sum_i   \, V_{ai} {M_1 M_i \over M_i^2 - t} V_{bi} \ ,
\eeqa{NMdefs}
with $a,b = L,R$.  If we can ignore CP violation, $N_{LR} = N_{RL}$, and,
similarly, $M_{LR} = M_{RL}$.
 The neutralino functions $M_{ab}(t)$ enter
the amplitudes for 
the $\ee$ and $\emem$ processes which require a helicity flip, and the 
functions $N_{ab}(t)$ enter the amplitudes for those processes which are 
helicity-conserving.

\begin{figure}
\begin{center}
\leavevmode
{\epsfxsize=3.20truein \epsfbox{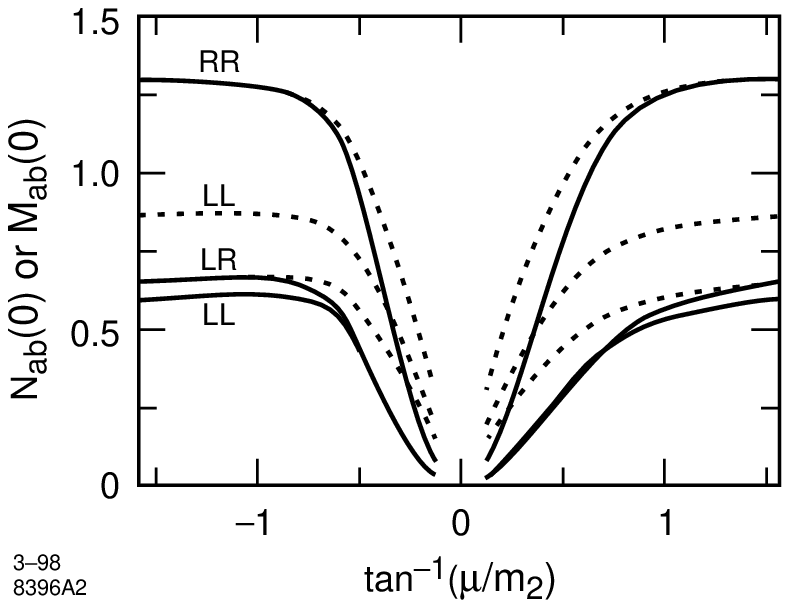}}
\end{center}
 \fcaption{The six neutralino functions, evaluated at $t=0$, 
displayed as  functions of 
       $\tan(\mu/m_2)$, for $\tan \beta=4$,
along the line in parameter space for which
   the mass of the lightest neutralino is 50 GeV.
    The functions $N_{ab}$ are drawn as solid lines, the $M_{ab}$ as 
   dotted lines.}
\label{NMfuncts}
\end{figure}

All of the neutralino functions must tend to zero in the Higgsino limit.
However, it turns out that they are still quite substantial for fairly
large
values of 
$m_2/|\mu|$.  The dependence of the six possible  functions on 
$m_2/\mu$ is shown in Figure \ref{NMfuncts}

The neutralino functions typically have a simple monotonic decrease with
$|t|$.  When we make use of them, we should think about extrapolating them 
to $t=0$.  In the gaugino limit, the values of these functions take  
simple values at $t=0$ for the cases $ab = RR$, $LR$:
\beq
    N_{RR}(0)\ , \ M_{RR}(0) \to {1\over c_w^2} \ \qquad \ 
  N_{LR}(0)\ , \ M_{LR}(0) \to {1\over 2c_w^2}  \ .
\eeq{RRlim}
To the extent that these predictions are not obeyed, the lightest neutralino
must have substantial Higgsino content.  On the other hand, the $t \to 0$ limit
of the  $LL$ 
neutralino functions varies with the ratio $m_1/m_2$:
\beq
  M_{LL}(0) \to {1\over 4c_w^2} + {1\over 4s_w^2}{m_1\over m_2} \qquad
  M_{LL}(0) \to {1\over 4c_w^2} + {1\over 4s_w^2}{m_1^2\over m_2^2} \ .
\eeq{LLlim}
The measurement of these functions then can be used to fix the value of
the ratio of gaugino masses.

\section{Cross Sections}

In writing the cross sections for selectron production, I will denote the 
superpartner of the right-handed electron by $\widetilde e_R$ and the 
superpartner of the left-handed electron by $\widetilde e_L$.  These 
particles of course are scalars and carry zero spin.  But the labels help
in tracking how the initial-state lepton helicity flows to the final-state
particles.  It is important to keep in mind, when discussing $\ee$ reactions,
that the $e^+_R$ goes with the $e^-_L$ and vice versa. 
In $\emem$ reactions, the helicity flow is transparent.  It is also important
to remember that, in most models, the the $\widetilde e_R$ and 
 $\widetilde e_L$ have masses which are substantially different. In the 
simplest models, the  $\widetilde e_L$ is heavier by a factor 1.3--2.5. 
So it is likely that the production of the two species of selectron can 
be distinguished kinematically.

\begin{figure}
\begin{center}
\leavevmode
{\epsfxsize=2.50truein \epsfbox{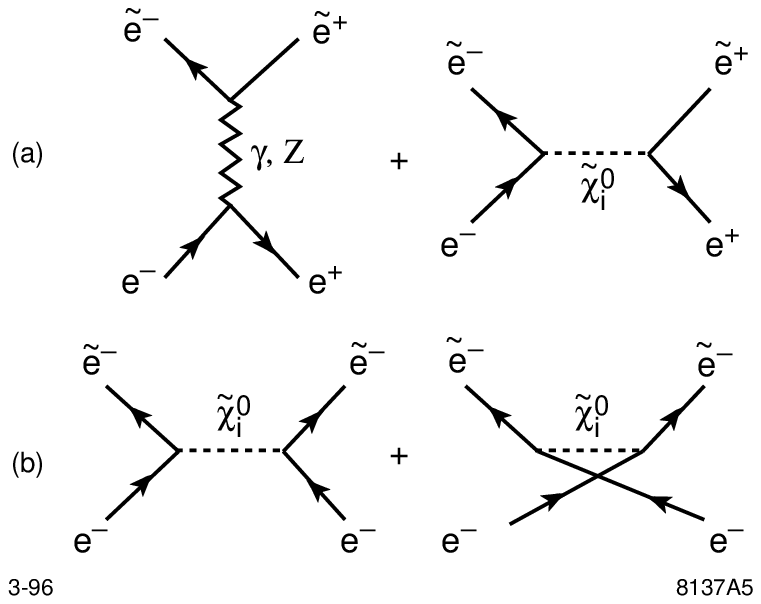}}
\end{center}
 \fcaption{Feynman diagrams contributing to the reactions 
(a) $\ee\to \se^+\se^-$, (b) $\emem \to \se^-\se^-$.  }
\label{efeyn}
\end{figure}

The Feynman diagrams for the reactions $\ee\to \se^+\se^-$ and 
$\emem \to \se^-\se^-$ are shown in Figure \ref{efeyn}.
Each possible final state couples to precisely one combination of initial
electron and electron-positron helicities, and also to one neutralino 
function.  Thus, each physically distinguished cross section is the 
square of a single helicity amplitude.  

I will now give  a list of formulae
for $d\sigma/d\cos\theta$ in the center of mass frame for each possible
reaction.   I denote the selectron velocity by $\beta$; for reactions with
unequal masses in the final state, $\beta = 2k/\sqrt{s}$, where $k$ is the 
common value of the final-state momentum.  Then we find
\beqa
\EREL \to \SE RR &:&  {\pi\alpha^2\over 2s}\beta^3\sin^2\theta 
    \bigl|{s\over M_1^2} N_{RR}(t) - (1 + {s_w^2\over c_w^2}{s\over s-\mz^2})
                    \bigr|^2  \CR
\EREL \to \SE LL &:&  {\pi\alpha^2\over 2s}\beta^3\sin^2\theta 
    \bigl|1 -  {\half - s_w^2\over c_w^2}{s\over s-\mz^2}
                    \bigr|^2  \CR
\ERER \to \SE RL &:&  {2\pi\alpha^2\over s}\beta {s\over M_1^2}
                    \bigl| M_{LR}(t) \bigr|^2  \CR
\ELEL \to \SE LR &:&  {2\pi\alpha^2\over s}\beta {s\over M_1^2}
                    \bigl| M_{LR}(t) \bigr|^2   \CR
\ELER \to \SE RR &:&  {\pi\alpha^2\over 2s}\beta^3\sin^2\theta
    \bigl|1 -  {\half - s_w^2\over c_w^2}{s\over s-\mz^2}
                    \bigr|^2  \CR
\ELER \to \SE LL &:&  {\pi\alpha^2\over 2s}\beta^3\sin^2\theta 
    \bigl|{s\over M_1^2} N_{LL}(t) -
 (1 + {(\half - s_w^2)^2\over c_w^2 s_w^2}{s\over s-\mz^2})
                    \bigr|^2   \CR
\EERR \to \SEE RR &:&  {2\pi\alpha^2\over s}\beta {s\over M_1^2}
                    \bigl| M_{RR}(t)+ M_{RR}(u) \bigr|^2  \CR
\EELR \to \SEE LR &:&  {\pi\alpha^2\over 2s}\beta^3\sin^2\theta 
    \bigl|{s\over M_1^2} N_{LR}(t) \bigr|^2 \CR
\EELL \to \SEE LL &:&  {2\pi\alpha^2\over s}\beta {s\over M_1^2}
                    \bigl| M_{LL}(t)+ M_{LL}(u) \bigr|^2\ . 
\eeqa{emrxs}
In formulae with identical particles in the final state, 
 $d\sigma/d\cos\theta$ is to be integrated over $0 \leq \cos\theta\leq 1$
only.
Note that, in amplitudes in which the $s$ channel contributions interfere
with the $t$-channel neutralino exchange, it is typically the neutralino
exchange that dominates.

\begin{figure}
\begin{center}
\leavevmode
{\epsfxsize=3.50truein \epsfbox{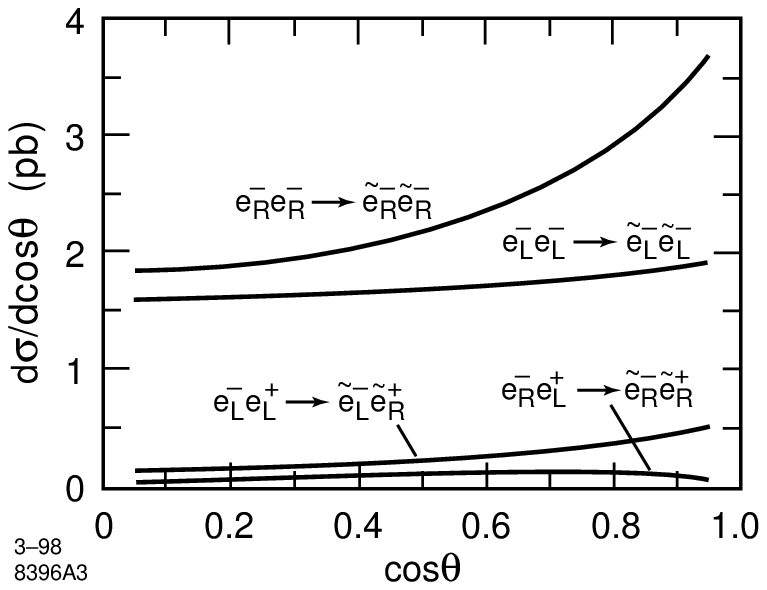}}
\end{center}
 \fcaption{Differential cross sections for four slepton production
 processes, computed at a point in the Higgsino region, with $m_2/\mu = -5$
 for $\tan\beta = 4$.
  I have taken $M_1 = 50$, $m(e_R) = 150$, $m(e_L) = 200$, $\sqrt{s} = 500$
    GeV.}
\label{css}
\end{figure}

From this table, we see that the reactions which produce $\se_R\se_R$ are
already useful for determining whether we are in the gaugino or Higgsino region
of the neutralino mixing problem.  The reactions which  produce
 $\se_L\se_L$ are the ones which are most sensitive to the ratio 
$m_1/m_2$.  For both of these final states, $\emem$ reactions have the 
interesting advantage that the amplitude near threshold is $s$-wave, and
so the cross section behaves like $\beta$ rather than $\beta^3$.
Depending on the kinematics, this difference can give a very substantial 
advantage in the size of the cross section to $\emem$.   In addition, 
the $\emem$ reactions have no destructive interference from the $s$-channel
diagrams.  Just as one
illustration of these effects,
 I plot in Figure \ref{css} the cross sections for the 
most important reactions at a point in the Higgsino region where
$m_2/\mu = -5$.

\section{Conclusions}

We have seen that the reactions $\ee \to \se^+\se^-$ and $\emem\to
\se^-\se^-$ have a wonderful formal simplicity.  By measuring these
cross sections, especially with polarized initial-state particles,
we have a direct
way to measure the parameters of neutralino mixing.  We have also seen
that $\emem$ has a certain advantage in these studies, both in the size
of cross sections and in the simplicity of the backgrounds which must be
controlled.

The simplicity of the $\emem$ reactions also give them an advantage in 
probing more sophisticated aspects of the electron-selectron coupling.
That is the subject of the presentations of Cheng, Feng, and Thomas
to these proceedings.

\nonumsection{Acknowledgements}

I am grateful to Clemens Heusch, Nora Rogers, and their colleagues at the 
University of California, Santa Cruz, for arranging this pleasant and fruitful 
conference.

\nonumsection{References}

\end{document}